\documentclass[prd,twocolumn,showkeys,amsmath,amssymb,nofootinbib,superscriptaddress]{revtex4}
\usepackage{graphicx}
\usepackage{epsfig}
\usepackage{bm}
\usepackage{amsfonts}
\usepackage[latin1]{inputenc}
\usepackage{amssymb}
\usepackage{color}
\usepackage{float}
\usepackage{amsmath}
\usepackage{dcolumn}
\usepackage{hyperref}

\def\be{\begin{equation}}
\def\ee{\end{equation}}
\def\bea{\begin{eqnarray}}
\def\eea{\end{eqnarray}}
\def\eqi{\begin{equation}}
\def\eqf{\end{equation}}
\def\eqia{\begin{eqnarray}}
\def\eqfa{\end{eqnarray}}

\newcommand{\CC}{\Lambda}

\newcommand{\mincir}{\raise
-3.truept\hbox{\rlap{\hbox{$\sim$}}\raise4.truept\hbox{$<$}\ }}
\newcommand{\magcir}{\raise
-3.truept\hbox{\rlap{\hbox{$\sim$}}\raise4.truept\hbox{$>$}\ }}

\newcommand{\rL}{\rho_{\Lambda}}

\newcommand{\ka}{\kappa}

\begin{document}

\begin{flushleft}
KCL-PH-TH/2019-{\bf 07}
\end{flushleft}

\title{Scalar Field Theory Description of the Running Vacuum Model: the Vacuumon}

\author{Spyros Basilakos$^{a,b}$, Nick E. Mavromatos$^{c}$ and Joan Sol\`a Peracaula$^d$}

\affiliation{$^a$Academy of Athens, Research Center for Astronomy and Applied Mathematics, Soranou Efessiou 4, 115 27 Athens, Greece. \\
$^b$ National Observatory of Athens, Lofos Nymfon,
11852, Athens, Greece. \\
$^c$Theoretical Particle Physics and Cosmology Group, Physics Department, King's College London, Strand, London WC2R 2LS, UK.\\
 $^{d}$Departament de F\'\i sica Qu\`antica i Astrof\'\i sica, and Institute of Cosmos Sciences (ICCUB), Univ. de Barcelona, Av. Diagonal 647 E-08028 Barcelona, Catalonia, Spain.}


\begin{abstract}
\vspace{0.2cm}
{We investigate the running vacuum model (RVM)} in the framework of scalar field theory.
This dynamical vacuum model provides an elegant global explanation
of the cosmic history, namely the universe starts from a non-singular initial
de Sitter vacuum stage, it  passes smoothly from an
early inflationary era  to a radiation epoch (``graceful exit") and finally
it enters the dark matter and dark energy (DE) dominated epochs, {where it can explain the large entropy problem and predicts a
mild dynamical evolution of the DE}.
Within this phenomenologically appealing context, we formulate
an effective {\it classical} scalar field description of the RVM through a field $\phi$,  called the {\it vacuumon},
which turns out to be very helpful for an understanding and practical implementation of the physical mechanisms of the running vacuum
during both the early universe and the late time cosmic acceleration. In the early universe the potential for the vacuumon
may be mapped to a potential that behaves similarly to that of the scalaron field of Starobinsky-type inflation at the {\it classical} level, whilst in the late universe it provides an effective scalar field description of DE. 
The two representations, however, are not physically equivalent since the mechanisms of inflation are entirely different. Moreover, unlike the scalaron, vacuumon is treated as a classical background field, and not a fully fledged quantum field, hence cosmological perturbations will be different between the two pictures of inflation.

\end{abstract}

\maketitle

\hyphenation{tho-rou-ghly in-te-gra-ting e-vol-ving con-si-de-ring
ta-king me-tho-do-lo-gy fi-gu-re}

\section{Introduction \label{sec:1}}
{The recent and past analyses of the Planck collaboration\,\cite{Planck2018,Planck2015,Planck2013}
provide significant  support to the general
cosmological paradigm}, namely that the observed Universe
is spatially flat and it is dominated by dark energy
(around $\sim 70\%$), while the rest corresponds to matter
(dark matter and baryons)
\cite{dataacc1,dataacc2,data1,data2,data3}.
From the dynamical viewpoint dark energy (hereafter DE) occupies
an eminent role
because it provides a theoretical framework towards explaining
the cosmic acceleration. The simplest candidate for DE is the
cosmological constant. The combination with
cold dark matter (CDM) and ordinary baryonic matter
builds the so-called $\Lambda$CDM model,
which fits extremely well the current
cosmological data. Despite its many virtues, the
$\Lambda$CDM model suffers from two fundamental problems, namely
the fine tuning problem and the cosmic
coincidence problem \cite{conpr1,conpr5,Pee2003,conpr2,JSPRev2013,conpr4}.
But it also suffers from some persistent phenomenological problems (called ``tensions'') of very practical nature, see e.g. \cite{conpr3} for a variety of them.  Perhaps the most worrisome ones nowadays are two.
One of them is. the mismatch between the CMB determination and the local value of the Hubble parameter\,\cite{Riess2016and2018}. Moreover, there is an unexplained  `$\sigma_8$-tension'\,\cite{Macaulay2013}) , which is revealed through the fact that the $\CC$CDM tends to provide higher values of that parameter than those obtained from large scale structure formation measurements.  Dynamical DE models seem to furnish a possible alleviation of such tensions, see e.g. \cite{SOLL17,SOLL18,JSP17,SolGoCruz18,Mehdi19}, and this is a very much motivating ingredient for focusing our attention on these kind of models. Let us note that these models can also appear as effective behavior of theories beyond General Relativity (GR). For instance, it has recently been shown\,\cite{BD-RVM} that Brans-Dicke gravity with a cosmological constant, when parameterized as a deviation of GR, can mimic dynamical dark energy, and more specifically  running vacuum (see below), and can provide a solution to those tensions, in particular to the acute $H_0$-tension (which is rendered harmless in this framework).

These problems have opened a window for plenty
cosmological scenarios which basically generalize the standard
Einstein-Hilbert action of GR by introducing
new fields in nature
(\cite{hor1,hor2,hor3,hor4}), or extensions of GR.
Among the large family of DE models,
very important in these kind of studies
is the framework of the
\textit{running vacuum model} (RVM)
~\cite{ShapSol,ShapSol2,ShapSolStef,SolStef05,Fossil07,ShapSol09} -- see
\cite{JSPRev2013,JSPRevs,SolGo2015} and references therein for a
detailed review. In fact the idea to have
a vacuum that depends on cosmic time (or redshift)
has a long history in cosmology and it is
perfectly allowed by the cosmological principle
(\cite{Ozer,bertolami,chenwu,lim00,lima7,many1,many2,aldrovandi,Schutz,Schutz2,Lima1,Lima2,Lima4,Lima5,SalimWaga,Waga,span1}).


In this context, the equation of state is similar to standard
$\Lambda$CDM $p_{\Lambda}=-\rho_{\Lambda}$, however the
vacuum density varies with time
$\rho_{\Lambda}=\rho_{\Lambda}(t)$.
The cosmological implications of various dynamical DE models
models have recently been analyzed both for the early
universe\,\cite{LBS2013,LBS2014,SolaGRF2015,LBS2016,BLS2013,Perico2013} as well as
for the late
universe\,\cite{GoSolBas2015,GoSol2015,Elahe2015} -- see also
\cite{BPS09,GrandeET11,FritzschSola2012,BasPolSol2012,BasSol2013,OldPerturb,Cristina}
for previous analyses and \cite{SOLL17,SOLL18,JSP17,SolGoCruz18,Mehdi19} for the most recent ones. In the context of the  RVM, the parameter
$\rho_\Lambda$ acts as a running parameter for the entire evolution\,\cite{SolGo2015} and evolves slowly as a power series
of the Hubble parameter. One finds that
the spacetime emerges from a non-singular initial de Sitter vacuum stage,
while the phase of the universe changes smoothly
from early inflation to a radiation era ("graceful exit").
After this period, the universe passes to the
dark-matter and $\Lambda$-dominated epochs before
finally entering a late-time de Sitter era.

In a previous work \cite{BasMavSol15}, we have provided a concrete examples of  a class of field theory models of the early Universe that
can be connected with RVM. Specifically, we have managed to reformulate the effective action of  Supergravity (SUGRA) inflationary models as an RVM, by
demonstrating that $\rho_{\Lambda}$ in such models can be written as
an even power series of the Hubble
rate, which can be naturally truncated at the $H^4$ term.
This is exactly the underlying framework
expected in the simplest class of
running vacuum models.
The RVM can also provide a clue for alleviating the fine tuning problem\,\cite{JSPRev2013}. Moreover, comparison of the RVM
against the latest cosmological data (SNIa+BAO+$H(z)$+LSS+BBN+CMB)
yield a quality fit that is significantly
better than the concordance $\Lambda$CDM, see\,\cite{SOLL17,SOLL18} for detailed analyses supporting this fact. A summary review is provided in \cite{JSPReview16,MG15}.
Interestingly, the RVM can be mimicked by Brans-Dicke theories as well,\cite{RVM-BD}.
Obviously, the aforementioned results have led to
growing interest in dynamical vacuum  cosmological models of the type $\Lambda=\Lambda(H)$
Therefore, there is every motivation
for further studying running vacuum models
from different perspectives, with the
aim of finding possible connections with fundamental aspects
of the cosmic evolution.
In point of fact, this is the main goal of the present article.
Specifically, we attempt to provide a scalar field description
of the RVM which is one of the most popular $\Lambda(H)$CDM models.
As we shall show, within this scalar field representation, the RVM can describe inflation and subsequent exit from it,
however the underlying physics is different from other approaches, like the Starobinsky approach.

The structure of the paper is as follows. The general framework of
the RVM is introduced in II. The basic theoretical
elements of the RVM are presented in sections III and IV.
The scalar field description of the RVM, based on the
{\it vacuumon} field is developed in section V.
Finally, our conclusions are summarized in section VI.

\section{Running vacuum cosmology} \label{sec:running}
{The general idea of dynamical vacuum
is a useful concept because it may provide and elegant global description
of the cosmic expansion. In this section, we briefly present the basic ingredients
of a specific realization of this idea which goes under the name of running vacuum model (RVM),
based on  an effective ``renormalization group (RG) approach'' of the cosmic evolution, see\,\cite{ShapSol,Fossil07} and \cite{JSPRev2013,SolGo2015}.
In fact, the RVM grants a unified dynamical picture for the
entire cosmic evolution\,\cite{LBS2013,BLS2013,Perico2013,LBS2014,SolGo2015,SolaGRF2015,LBS2016}.} It
connects smoothly the first two cosmic eras,
inflation and standard
Fridman-Lema\^{\i}tre-Robertson-Walker (FLRW) radiation,
and subsequently allows the universe to naturally enter the matter
and dark energy domination in
the present time, in which it still carries a mild dynamical
behavior compatible with the
current observations\,\cite{GoSolBas2015,GoSol2015}.



{In the old decaying vacuum models mentioned above (see e.g. \cite{Overduin98} for a review ) the
time-dependent cosmological term  was usually some ad hoc function
$\CC(t)$ of the cosmic time or of the scale factor.  However,  the
running vacuum idea is inspired in the context of QFT in curved spacetime. It is based on
the effective RG approach to the evolution of the universe, and hence on the existence of a
running scale, $\mu_c=\mu_c(t)$, which is connected to the dynamical
variables of the cosmic evolution.  Such variable is not known a priori. In particle
physics, for example, one associates it to the characteristic energy scale of the process
in an accelerator, or to the mass of the decaying particle in the proper reference frame. In cosmology, however, it is
more difficult, but a natural ansatz within the FLRW metric is to associate the RG running scale to the Hubble parameter,
 i.e. to the expansion rate of the cosmic evolution, $\mu_c(t)=H(t)$ -- see \cite{JSPRev2013,SolGo2015} and references therein.}
Following the notations in these references,
the RG equation takes the general form:
\begin{equation}\label{runningrho}
\frac{d\, \rho_\Lambda (t)}{d\, {\rm ln}H^2 } = \frac{1}{(4\pi)^2}
\sum_i \Big[a_i M_i^2 H^2 + b_i H^4 + c_i \frac{H^6}{M_i^2} + \dots
\Big],
\end{equation}
Let us emphasize that  $\rho_{\Lambda}=\Lambda/\kappa^2$ in this equation is treated as the vacuum energy density, due to the presence of $\Lambda(t)$, with pressure being given by the following equation of state (EoS)
\begin{equation}\label{eosrvm}
p_{\Lambda}(t)=-\rho_{\Lambda}(t).
\end{equation}
We would like to stress that
the latter EoS does not depend on whether the vacuum is dynamical or
not.

In general $\mu_c^2$ can be associated to a linear combination of
$H^2$  and $\dot{H}$ and the variety of terms appearing on the
r.h.s. of (\ref{runningrho}) can be richer\,\cite{SolGo2015}, but
the canonical possibility is the previous one and hereafter we
restrict to it. As shown in specific analyses\cite{SOLL17,SOLL18}, there are no dramatic differences in the phenomenological results obtained in the presence of the additional $\dot{H}$ term. Furthermore, these terms are not essential either for the description of the early universe, at least  within the RVM since  $\dot{H}\simeq 0$ during inflation, see \cite{LBS2013,BLS2013,Perico2013}.

The coefficients $a_i, b_i, c_i \dots$ in (\ref{runningrho}) are dimensionless
and receive contributions from
loop corrections of boson and fermion matter fields with different
masses $M_i$.
The expression (\ref{runningrho})
has to be understood as a general
ansatz for the vacuum energy density in an expanding universe.
The reason for choosing even powers of $H$ is due to
general covariance of the effective action of QFT
in a curved background. 
Although we cannot presently
quote its precise form, it has been found
\cite{ShapSol,ShapSol2,ShapSolStef,SolStef05,Fossil07,ShapSol09} that it can only include
even powers which can emerge from the contractions
of the metric tensor with the derivatives of the scale
factor. Moreover, the current vacuum models have recently been linked with
a potential variation of the so-called fundamental
constants of nature \cite{FritzschSola2012}.


Notice that if the evolution of the Universe is
restricted to eras close 
the Grand Unified Theory (GUT) scale, then
for all practical purposes it is at most
the $H^4$ terms (those with dimensionless coefficients $b_i$) that can
contribute significantly.
There is no possible inflation without the
higher order powers of $H^n\ (n>2)$, as
explained in detail in \cite{LBS2013,BLS2013,Perico2013}.

The $H^2$ term is of course negligible at
this point, and the higher powers of $H^n$ for $n=6,8,..$ are
suppressed by the corresponding inverse powers of the heavy masses $M_i$, which go to the denominator, as required by the decoupling theorem. In the
scenarios of dynamical breaking of local supergravity discussed in
Ref.~\cite{ahm,ahmstaro,emdyno}, the breaking and the associated
inflationary scenarios could occur around the GUT scale, in
agreement with the inflationary phenomenology suggested by the Planck
data~\cite{Planck}, provided Jordan-frame supergravity models
(with broken conformal symmetry) are used, in which the conformal
frame function
acquired, via appropriate dynamics, some non
trivial vacuum expectation value\,\cite{BasMavSol15}.
For these situations, therefore,
corrections in (\ref{runningrho}) involving higher powers than $H^4$
will be ignored.

Remarkably, it can be shown that the presence of the term  $H^4$ in the effective expression of the vacuum energy density can be the generic result of the low-energy effective action based on the bosonic gravitational multiplet of string theory, see \cite{Anomaly2019a}. The unavoidable presence of the CP-violating gravitational Chern-Simons term associated with that action turns out to lead to an effective $\sim H^4$ behavior when averaged over the inflationary spacetime, in the presence of primordial gravitational waves. This higher order term triggers inflation within the context of the RVM, see Sec.\ref{deSiter}. It follows that the
entire history of the universe can be described in an effective RVM language upon starting from the fundamental massless bosonic gravitational multiplet of a generic string theory. The reader is invited to study \cite{Anomaly2019a} for details and \cite{GRF2019} for a summary of the underlying framework.

In the next sections, we study the running vacuum model by integrating
(\ref{runningrho}), following the approach of
\,\cite{LBS2013,BLS2013,Perico2013,LBS2014}.
Before doing so, let us briefly review the main ingredients of RVM.

\vspace{0.2cm}
\section{Dynamical vacuum and running vacuum model}
 \label{sec:4}

The aim of this section is to demonstrate that there exists a
{family of time-dependent effective dynamical vacuum models} of
{\em running type}, i.e. the class of the running vacuum models
(RVM's) introduced in  Sect.\,\ref{sec:running}, which characterize
the evolution of the Universe from the exit of the Starobinsky
inflationary phase till the present era.

Following the notations of
\cite{ShapSol,LBS2013,BLS2013,Perico2013,LBS2014}
Eq.\,(\ref{runningrho}) is approximated by
\begin{equation}\label{runningrho2}
\frac{d\, \rho_\Lambda (t)}{d\, {\rm ln}H^2 } = \frac{1}{(4\pi)^2}
\sum_i \Big[a_i M_i^2 H^2 + b_i H^4 \Big].
\end{equation}
Performing the integration we find
\begin{equation}\label{lambda}
\rho_{\Lambda}(H) = \frac{\Lambda(H)}{\kappa^2}=
\frac{3}{\kappa^2}\left(c_0 + \nu H^{2} + \alpha
\frac{H^{4}}{H_{I}^{2}}\right) \;,
\end{equation}
where $\kappa^{2}=8\pi G$,
$c_0$ is an integration constant (with dimension $+2$ in
natural units, i.e. energy squared) which can be constrained
from the cosmological data \cite{BPS09,GoSolBas2015}.
Also, the dimensionless coefficients are given by
\begin{equation}\label{eq:nuloopcoeff}
\nu=\frac{1}{48\pi^2}\, \sum_{i=F,B} a_i\frac{M_i^2}{M_{\rm
Pl}^2}\,,
\end{equation}
and
\begin{equation}\label{eq:alphaloopcoeff}
\alpha=\frac{1}{96\pi^2}\, \frac{H_I^2}{M_{\rm Pl}^2}\sum_{i=F,B}
b_i\,.
\end{equation}

For the present universe, $H=H_0\sim 10^{-42}$ GeV,  it is obvious that the $\sim H^4$ contribution can be completely neglected as
compared to the $\sim H^2$ component of the vacuum energy density (\ref{lambda}).  However, for the early universe the value
of $H$ becomes much larger and the $\sim H^4$  contribution is dominant.  We shall confirm this fact quantitatively in the next
section in the light of the solution of the cosmological equations.

Practically, the coefficient $\nu$ can be seen as the
reduced (dimensionless) beta-function for the RG running of
$\rho_{\Lambda}$ at low energies, while $\alpha$ has a similar
behavior at high energies. Notice that the index $i$ depends on whether
bosons ($B$) or fermions ($F$) dominate in the loop contributions.
Of course, since the dimensionless
coefficients $(\nu,\alpha)$ play the role of
one-loop beta-functions (at the respective low and high energy
scales) they are expected to be naturally small because $M_i^2\ll
M_{\rm Pl}^2$ for all the particles, even for the heavy fields of a
typical GUT. Indeed, within GUT $\nu$ lies
in the range $|\nu|=10^{-6}-10^{-3}$ \cite{Fossil07}, while
$\alpha$ is also small ($|\alpha|\ll 1$),
because the scale of inflation $H_I$ is certainly below the Planck
scale.
The latter predictions are in agreement with the observational
constraints.   In point of fact, from  the joint likelihood
analysis of various cosmological data
(supernovae type Ia data, the CMB shift
parameter, and the Baryonic Acoustic Oscillations), it has been found
$|\nu|={\cal
O}(10^{-3})$\,\cite{BPS09,GrandeET11,GoSolBas2015,Elahe2015,GoSol2015}
(see also \cite{SOLL17,SOLL18,AdriaJoan2017}).


Let us finally clarify the possible effect of the $\dot{H}$ terms, which we mentioned briefly in the previous section.
Since  ${\dot H}=(1+q)H^{2}$ (with $q$ the deceleration parameter), it can been shown that the effect of that term
in the measured
cosmological observables can be taken
into account by ``renormalizing'' the effect of the $H^2$ terms.
Moreover, during inflation we have $H\simeq$ constant and as previously indicated ${\dot H}$ can be neglected. Therefore, terms of the form  $\dot{H}^2$  and $\dot{H} H^2$  are negligible
and inflation is dominated by the single term proportional to $H^4$ in the context of the RVM.

\section{Running Vacuum versus Scalar field:  the  vacuumon}

Let us start here
with the Einstein-Hilbert action
\begin{equation}
S_{R,\Lambda}=\int d^{4}x\sqrt{-g}\left[
\frac{1}{2\kappa^{2}}(R-2\Lambda)
+\mathcal{L}_{m}\right]\,,  \label{action1}%
\end{equation}
where $\kappa=\sqrt{8\pi G}$, with $G$ the four-dimensional Newton constant,
 $\rho_{\Lambda}(t)=\Lambda (t)/\kappa^2$
and $\mathcal{L}_{m}$
is the Lagrangian of matter.
Notice that the cosmological equations are expected
to be formally equivalent to
the standard $\Lambda$CDM case, due to the Cosmological Principle which is
embedded in the FLRW metric. The latter is perfectly compatible with the possibility of a
time-evolving cosmological term ~\cite{ncstrings}.
If we vary the action (\ref{action1}) with
respect to the metric we obtain the field equations
\begin{equation}
R_{\mu \nu }-\frac{1}{2}g_{\mu \nu }R=\kappa^2\,  \tilde{T}_{\mu\nu}\,,
\label{EE}
\end{equation}
where $\tilde{T}_{\mu\nu}\equiv
p_{\rm tot}\,g_{\mu\nu}+(\rho_{\rm tot}+p_{\rm tot})\,U_{\mu}U_{\nu}$,
is the total energy momentum tensor, with
with $\rho_{\rm
tot}=\rho_{m}+\rho_{\Lambda}$ and $p_{\rm
tot}=p_{m}+p_{\Lambda}=p_{m}-\rho_{\Lambda}$.
As already mentioned, $\rho_{\Lambda}=\Lambda(t)/\kappa^2$ is the vacuum energy
density related to the presence of $\Lambda(t)$ with EoS \eqref{eosrvm}.
The quantity
$\rho_{m}$ is the density of matter-radiation and
$p_{m}=\omega_{m} \rho_{m}$ is the corresponding pressure, where
$\omega_m$ is the EoS parameter:  $w_{m}=0$ for nonrelativistic matter and
$w_{m}=1/3$ for relativistic (i.e. for the radiation component).

In the framework, of a spatially flat FLRW spacetime, we
obtain the Friedmann equations in the presence of a running
$\Lambda$-term: \be
 \kappa^{2}\rho_{\rm tot}=\kappa^2 \rho_{m}+\Lambda = 3H^2 \;,
\label{friedr} \ee
\be
\kappa^{2}p_{\rm tot}=\kappa^2 p_{m}-\Lambda =-2{\dot H}-3H^2
\label{friedr2} \ee and the Ricci scalar
\begin{equation}
\label{SF.3b} R=g^{\mu\nu}R_{\mu\nu}= 6(2H^{2}+\dot{H}) \;,
\end{equation}
where the overdot denotes derivative with respect to cosmic time
$t$. Following standard lines the Bianchi identities
$\bigtriangledown^{\mu}\,{\tilde{T}}_{\mu\nu}=0$ for $G=$const.
reduce
\begin{equation}
\dot{\rho}_{m}+3(1+\omega_{m})H\rho_{m}=-\dot{\rho_{\Lambda}}\,.
\label{frie33}
\end{equation}
where one may check that there is an exchange between matter and
vacuum.

Using equations (\ref{friedr}), (\ref{friedr2}) and
(\ref{frie33}), we derive the main differential equation that
governs the cosmic expansion, namely
\begin{equation}
\dot{H}+\frac{3}{2}(1+\omega_{m})
H^{2}=\frac12\,\kappa^2(1+\omega_{m})\rho_{\Lambda}=
\frac{(1+\omega_{m})\Lambda}{2}\,. \label{frie34}
\end{equation}
Inserting Eq.(\ref{lambda}) into the above differential equation
and solving it, one finds
\be
\label{HE} \dot
{\dot H}+\frac{3}{2}(1+\omega_m)H^2
\left(1-\nu-\frac{c_0}{H^2}-
\alpha\frac{H^2}{H_{I}^2}\right)=0 .
\ee
Below, we discuss the cosmic history and
the scalar field description of RVM.

Before embarking onto that, we feel commenting on an important feature of the RVM approach, namely the fact that
terms of order ${\mathcal O}(H^4)$ and higher, that appear in the evolution \eqref{HE} of the Hubble parameter, owe their existence to the corresponding higher-order terms of the
RG-like vacuum energy density $\rho_\Lambda$ \eqref{lambda}. Such terms might also be associated with higher curvature terms in the effective action, in models beyond General Relativity, such as strings, although there might be subtleties in such an association, as we shall discuss in the next subsection, within the context of the Starobinsky model of inflation~\cite{staro}.

Indeed, on the right-hand side of \eqref{EE} we kept the ``Running Vacuum'' term
in an effective description, via $\Lambda(t)$, without specifying its microscopic origin. In general, one might have higher curvature terms appearing in the tensor $\tilde T_{\mu\nu}$, due, e.g. to quadratic curvature terms of the type appearing in the Starobinksy model~\cite{staro}, induced by conformal anomalies, or, in the context of string-inspired effective actions, by non-constant-dilaton $\Phi (x)$ terms-coupled to Gauss-Bonnet quadratic curvature invariants, $ \int d^4 x\sqrt{-g}\, e^{-2\Phi(x)} (R_{\mu\nu\rho\sigma} R^{\mu\nu\rho\sigma} - 4R_{\mu\nu} R^{\mu\nu} + R^2)$ or gravitational axion $b(x)$ terms, originating in four space-time dimensions from the antisymmetric tensor field of the massless bosonic string multiplet,  coupled to gravitational Chern-Simons anomalous terms in a CP-violating manner, $\int d^4 x \sqrt{-g}\, b(x) R_{\mu\nu\rho\sigma} \tilde R^{\mu\nu\rho\sigma}$, where the tilde over the Riemann tensor denotes the dual $\tilde R^{\mu\nu\rho\sigma} = \frac{1}{\sqrt{-g}}\,  \epsilon^{\mu\nu\alpha\beta} R_{\alpha\beta}^{\,\,\,\,\,\,\,\,\rho\sigma}$, where $\epsilon_{\mu\nu\rho\sigma}$ is the Minkowski-flat space-time totally antisymmetric symbol, $\epsilon^{0123}=+1$ etc. Such higher order terms also lead to contributions to the energy density of the Universe of ``running vacuum type'', in particular $H^4$ terms.

This is rather remarkable and has been demonstrated explicitly in \cite{Anomaly2019a,GRF2019} for the gravitational Chern Simons terms, in the presence of primordial gravitational waves, which induce condensates of the Chern-Simons terms that lead to de Sitter space time, but it can be extended to include all other higher curvature terms. Nonetheless, the presence of extra scalar modes like dilaton and/or the Starobinsky-model scalaron can be distinguished from the scalar mode encoded in the RVM contributions (``vacuumon''), as we shall discuss below, although their co-existence in certain models cannot be excluded. In general, the cosmological perturbation effects of scalaron and ``vacuumon'' models can be different.

In fact, as we shall see, the vacuumon is only viewed as a {\it classical background field}, useful for a unified description of the evolution of the Universe from a de Sitter (inflationary) phase to the present era. The reader should recall that the RVM, which the vacuumon describes in scalar field language, is based on a vacuum energy density function interpolating continuously the various epochs, {\it cf.} Eq.\,\eqref{lambda}. As such, it should be stressed, that this is a very different view as compared to more standard approaches, including Starobinsky inflation. In the latter, for instance, the scalaron appears as a physical particle associated to fluctuations of a quantum field. This is in stark contrast with the classical vacuumon picture since in the RVM there is no reheating caused by the decay of an intermediate state consisting of massive particles, but rather a fast and continuous process of heating up of the early universe whereby the inflationary vacuum decays into radiation and smoothly gives rise to the radiation phase\,\,\cite{LBS2013,LBS2014,SolaGRF2015,LBS2016,BLS2013,Perico2013}.

It is thus of foremost importance to realize that our framework does not aim at a microscopic interpretation. In fact, we remain at the level of the classical fields and ultimately rely on a thermodynamical description well along the line of the mentioned papers.  In this respect, let us mention that the entropy problem\,\cite{KolbTurner} in this kind of models is nonexistent and the cosmic evolution can be shown to be perfectly consistent with the generalized second law of thermodynamics\,\cite{SolaGRF2015}.
One of the aims of the present work is to discuss in detail such issues once the above fundamental distinction between the microscopic scalaron (or, in general, inflaton like)  picture and the classical vacuumon picture has been clearly established.

\subsection{Early de Sitter - radiation: Scalar field description}\label{deSiter}
Initially, from Eq.(\ref{HE}) we easily identify
inflation (de Sitter phase), namely
there is the constant value solution\footnote{A study concerning
the self-consistent de Sitter stage can be found in \cite{Dowker1976}.}
$H^2=(1-\nu)H_I^2/\alpha$ of Eq.(\ref{HE}), which is valid in the
early universe for which $c_0/H^2\ll 1$. It should be now clear that in the absence of the $\sim H^4$ term in Eq.(\ref{lambda}) such de Sitter solution would not exist. Therefore in order to trigger inflation it is necessary to go beyond the $H^2$ contribution to the vacuum energy-density, although not all higher powers are on equal footing. As explained, covariance requires the participation of an even number of times derivatives of the scale factor, and we have adopted the simplest available possibility, $\sim H^4$.
As already remarked, the presence of the  $H^4$-term in the effective expression of the vacuum energy density (\ref{lambda}) can actually be the generic result of gravitational Chern Simons
terms (averaged over the inflationary spacetime) that characterise the string-inspired low-energy effective action based on the bosonic gravitational multiplet of string theory\,\cite{Anomaly2019a}. Therefore, the RVM description
we are considering here can be associated with the effective treatment of such large class of well-motivated fundamental theories.

The inflationary solution can then be made explicit upon integration of Eq.(\ref{HE}) when $c_0/H^2\ll 1$ holds. We find
\begin{equation}\label{HS1}
 H(a)=\left(\frac{1-\nu}{\alpha}\right)^{1/2}\,\frac{H_{I}}{\sqrt{D\,a^{3(1-\nu)(1+\omega_m)}+1}}\,,
\end{equation}
where $D>0$ is the integration constant. For the early
universe we consider that matter is essentially relativistic, hence here we
impose $\omega_{r}\equiv \omega_m=1/3$, hence $p_{r}=\rho_{r}/3$.
Interestingly, we observe from (\ref{HS1}) that in the case of
$Da^{4(1-\nu)} \ll 1$ the universe starts from an unstable early de Sitter phase
[inflationary era,
$H^2=(1-\nu) H_I^2/\alpha$] dominated by the huge value $H_I$
which is potentially related to the GUT scale.
Notice that within this regime the ratio between the $\sim H^4$  and the $\sim H^2$ terms in Eq.\,(\ref{lambda}) is
$(\alpha/\nu)\,H^2/H_I^2\simeq  (\alpha/\nu)\,(1-\nu)/\alpha\simeq 1/\nu\gg1$  (since $|\nu|\ll 1$) and therefore the  $\sim H^4$ term
is dominant over  the $\sim H^2$ one. This means that in this regime we can neglect $\nu$. This confirms quantitatively the statement
made in Section III.
On the other hand, after the early de Sitter epoch, in particular
for $Da^{4(1-\nu)} \gg 1$, the universe enters
the standard radiation phase.
This behavior is confirmed
from the form of $\rho_{\Lambda}$ and $\rho_{r}$.
Upon neglecting the terms
$\nu$ and $c_0/H^2$ in this early epoch,
which is fully justified, we insert (\ref{HS1}) into (\ref{lambda})
and obtain:
\begin{equation}\label{eq:rLa}
  \rho_\Lambda(a)=\frac{{\rho}_I}{\alpha}\,\frac{1}{\left(1+D\,a^{4}\right)^{2}}\,.
\end{equation}
Substituting the vacuum density into Eq.(\ref{frie33}) and solving this differential equation we
obtain the following solution:
\begin{equation}\label{eq:rhor}
 \rho_r(a)=\frac{{\rho}_I}{\alpha}\,\frac{D\,a^{4}}{\left(1+D\,a^{4}\right)^{2}},
\end{equation}
where $\rho_I=3H_I^2/\kappa^2$ is the critical density during inflation.
Obviously, the aforementioned expressions tell us that
there is as absence of singularity in the initial state, namely the Universe
starts at $a=0$ with a huge vacuum energy density $\rho_I/\alpha$
(and $\rho_{r}=0$) which is gradually transformed
into relativistic matter.
Asymptotically, we recover the usual behavior
$\rho_r \sim  a^{-4}$, while
the vacuum energy density $\rho_\Lambda\sim a^{-8}\ll\rho_r$
becomes essentially negligible.
That is, graceful exit is achieved.

Despite the fact that
the origin of the RVM is based on
the effective action of
QFT in curved space-time, the form the action
is not known in general\,\cite{ShapSol09}; at present such a task has only been
achieved in specific cases~\cite{Fossil07}.
However, using a field theoretical language,
via an effective {\it classical} scalar field $\phi$, it is possible to
write down the basic field equations for the RVM~\cite{BasMavSol15}.
In the present work, we term the field $\phi$ the {\it early vacuumon}\footnote{{In the past, a scalar field called the {\it cosmon}\,\cite{PSW}
was proposed to solve the old cosmological constant problem\,\cite{conpr1}. The {\it vacuumon}, instead, does a different function, it is not intended for solving the CC fine tuning problem but rather to describe the running vacuum in terms of a scalar effective action. Therefore, it can help alleviate the cosmic coincidence problem, and in general it shares benefits with quintessence models, with the tacit understanding, however, that it just mimics the original RVM, whose fundamental action is not  known, except in some specific cases, as mentioned above.}}.
Combining Friedmann's Eqs.(\ref{friedr})-(\ref{friedr2}) and
following the standard approach
$\rho_{\rm tot}\equiv \rho_{\phi}={\dot \phi}^{2}/2+U(\phi)$
and $p_{\rm tot}\equiv p_{\phi}={\dot \phi}^{2}/2-U(\phi)$ we find
\begin{equation}
\dot{\phi}^{2} =-\frac{2}{\kappa^{2}}\dot{H} \;, \label{ff3}
\end{equation}
\begin{eqnarray}
\label{Vz} U&=&\frac{3H^{2}}{\kappa^{2}}\left(
1+\frac{\dot{H}}{3H^{2}}\right)= \frac{3H^{2}}{\kappa^{2}}\left(
1+\frac{aH^{'}}{3H}\right)\nonumber\\
&=&\frac{3H^{2}}{\kappa^{2}}\left(
1+\frac{a}{6H^2}\,\frac{d H^2}{da}\right) \;,
\end{eqnarray}
where $U(\phi)$ is the effective potential. Notice that
$\dot{H}=aHH^{'}$ with prime denoting derivative with
respect to the scale factor. Integrating Eq.(\ref{ff3}) we obtain
\begin{equation}
\label{ppz} \phi= \pm \int \left( -\frac{2\dot{H}}{\kappa^{2}}\right)^{1/2}
dt = \pm \frac{\sqrt{2}}{\kappa}\int
\left(-\frac{H^{'}}{aH}\right)^{1/2}da\;.
\end{equation}
Here we postulate the positive sign, while as we will discuss below
in the late era we use the minus.

Now, for $\omega_m=1/3$ the Hubble parameter (\ref{HS1}) becomes
\begin{equation}\label{HS11}
 H(a)=\left(\frac{1}{\alpha}\right)^{1/2}\,
\frac{H_{I}}{\sqrt{D\,a^{4}+1}}\,.
\end{equation}
We would like to point out that
we have imposed $\nu=0$ in Eq.(\ref{HS1}), which has no effect
for the study of the early universe.

Within this framework, integrating Eq.(\ref{ppz}) in the range $[0,a]$
with the aid of
Eq.(\ref{HS11})
we arrive at

\begin{eqnarray}\label{ppz1}\nonumber
\phi(a) & = & \frac{1}{\kappa}\;{\rm sinh}^{-1}\left( \sqrt{D}a^{2}\right),\\
& = & \frac{1}{\kappa}{\ln}\left(
\sqrt{D}a^{2}+\sqrt{Da^{4}+1}\right)\;.
\end{eqnarray}

\begin{figure}
\includegraphics[width=0.45\textwidth]{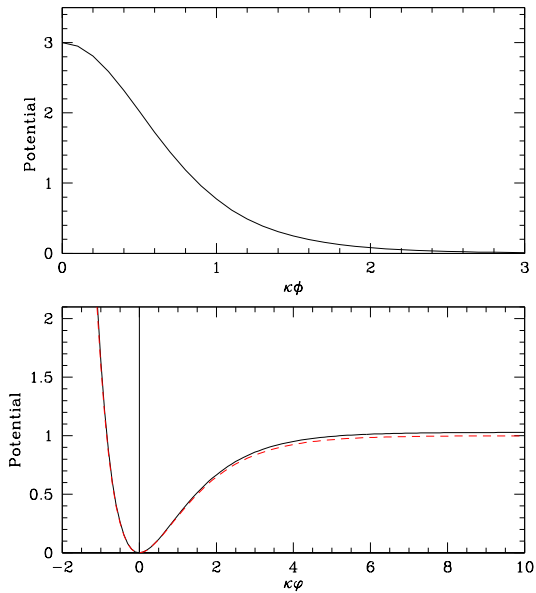}
\caption{{\it Upper panel}:The vacuumon classical
effective potential \eqref{Pott} versus the vacuumon field $\kappa \phi$.
The decay of the RVM vacuum is clearly seen by the shape of this
potential as the classical vacuumon field increases.
The radiation and matter dominated eras occur for some large
(but finite) value of $\kappa \phi$.
{\it Lower panel:}
The RVM effective potential $\alpha \kappa^{2} U/H^{2}_{I}$ (solid line) versus
the {\it scalaron} field $\kappa \varphi$. The dashed line {corresponds to  $4\kappa^2/({3\cal M}^2)$ times the Starobinsky
effective potential},  Eq.(\ref{staropotent}). The regime of validity of the comparison between the models is the positive axis on both fields, indicated
by the perpendicular line.
The reader should bear in mind, however, that the apparent mapping
in $\kappa \varphi$-space
between the two potentials does not imply a similarity of the underlying mechanisms for inflation, which are entirely different. Moreover, when expressed in terms of the scalaron, the kinetic term of the vacuumon would look non canonical}.
\label{fig:potstar2}
\end{figure}

Also using Eqs.(\ref{Vz})-(\ref{HS11})
the effective potential is written as
\begin{equation}
U(a)=\frac{H^{2}_{I}}{\alpha \kappa^{2}}\;\frac{3+Da^{4}}{(1+Da^{4})^{2}},
\end{equation}
and thus
\begin{equation}
\label{Pott}
U(\phi)=\frac{H^{2}_{I}}{\alpha \kappa^{2}}\; \frac{2+{\rm
cosh}^{2}(\kappa \phi)} {{\rm cosh}^{4}(\kappa \phi)}  \;.
\end{equation}
It should be stressed once more that the
early vacuumon field is classical, hence there is no issue in attempting to
considering the effects of quantum fluctuations of $\phi$ on the ``hill-top'' potential \eqref{Pott}, depicted in fig.~\ref{fig:potstar2}.
In fact, if $\phi$ were a fully fledged quantum field, such as the
conventional inflation (which is not the case here), the potential
\eqref{Pott}
provides slow-roll parameters which
fit at $2\sigma$ level the optimal range indicated by the Planck
cosmological data on single-field inflation~\cite{Planck}.

We would like to finish this section with a brief discussion regarding
the recent Planck results.
Specifically the results provided by the Planck team ~\cite{Planck}
have placed tight restrictions on
single scalar-field models of slow-roll inflation, supporting
basically models with very low tensor-to-scalar fluctuation ratio $
r = n_T/n_s \ll 1$, with a scalar spectral index $n_s \simeq 0.96 $
and no appreciable running. The upper bound found by Planck
team~\cite{Planck} on this ratio, is $r < 0.10$, but their favored regions
point towards $r \le 10^{-3}$. These results are in agreement
with the predictions of the Starobinsky-type (or
$R^2$-inflation, where $R$ is the Ricci scalar)
models of inflation ~\cite{staro,staro2,Vilenkin}.
For this inflationary paradigm the action is
\begin{equation}\label{staroaction}
S=\int d^4 x \sqrt{-g}\,  \left( R  + \beta  \, R^2 \right)~,  \quad \beta > 0~,
\end{equation}
while
the effective
potential $V_{\rm eff}(\varphi)$ is given by:
\begin{eqnarray}\label{staropotent}
 V_{\rm eff}(\varphi ) = \frac{3{\cal M}^2\left( 1 - e^{-\sqrt{\frac{2}{3}} \, \ka\varphi } \right)^2}{4\,\ka^2},
\end{eqnarray}
where $\varphi$ is the {\it scalaron} field, {characteristic of Starobinsky's inflation\,\cite{staro}}.
One may check that the {\it scalaron} mass, which can be viewed as
the new gravitational
degree of freedom that the conformal transformation was able to
elucidate from the Starobinsky action, is indeed provided by
parameter ${\cal M}=\sqrt{8\pi/3\beta}$, where
$\left.\frac{d^2 V_{\rm eff}(\varphi)}{d\varphi^2}\right|_{\varphi=0}={\cal
M}^2$.

Since the Starobinsky inflationary model
fits extremely well the Planck data on inflation~\cite{Planck} we would like
to combine it with that of the RVM model.
In other words, we are interested to see if there is a connection
between {\it early vacuumon} and {\it scalaron} fields for which the
RVM potential equals  that of Starobinsky.
A related formulation to Starobinsky framework is  anomaly-induced inflation, which has its own merits for describing inflation and graceful exit\,\cite{ShapSol3,ShapSol4}. One can actually find also a connection with the RVM in this formulation based on a nonlocal effective action generated by the conformal anomaly\,\cite{Fossil07}. A comparison between the two inflationary frameworks can be found in \cite{SolGo2015}.

However, we should stress that such an equality is purely {\it formal}.  The two models correspond to different effective actions\,\cite{BasMavSol15}.  As a concrete example let us consider the RVM in the early de Sitter phase of the Universe. As already mentioned, in such a case the underlying physics is well described by making the approximation that the dominant term in the RVM energy density $\rho_\Lambda (H)$ is the quartic $\propto H^4$, i.e. we may set $c_0,\nu =0$ in \eqref{lambda}. This implies that the RVM effective action can be approximated by
\begin{align}\label{steps1}
 S_{R,\Lambda}& =  \int d^4 x \sqrt{-g}\,  \left[\frac{1}{2\kappa^{2}} R -\rL(H)\right] \nonumber \\
    &\sim \frac{1}{2\kappa^{2}}\int d^4 x \sqrt{-g}\,  \left(R-6\alpha\frac{H^{4}}{H^{2}_{I}} \right)~.
\end{align}
Upon replacing $H^4$ by the square of the Ricci scalar, which is to be expected during epochs where $H$ is approximately constant, such as the inflationary (de Sitter) era, one may then write
\begin{equation}\label{steps2}
  S_{R,\Lambda} \simeq \int d^4 x \sqrt{-g}\,  \frac{1}{2\kappa^{2}} \Big(R - \alpha\frac{R^{2}}{24\, H^{2}_{I}}\Big)\,.
\end{equation}
Notice that, since $\alpha > 0$ in our case, the RVM model is \emph{not equivalent} to a Starobinsky-type model, for which the effective Lagrangian has the form (\ref{staroaction}), corresponding to a negative $\alpha$ coefficient in (\ref{steps1}).
The reason lies in the fact that the metric tensors between the two models, (\ref{staroaction}) and (\ref{steps1}), are different, related by a non-trivial  conformal transformation involving the scalar field\,\cite{BasMavSol15}. Nonetheless, we can always rewrite the potential of the RVM as a ``Starobinsky-like potential'' via the transformation \eqref{XXXa}, \eqref{XXX}, without reference to the microscopic Starobinsky higher curvature model.  The two models are, therefore, equivalent only at the level of effective potentials, and for this reason the vacuumon can share the same successful description of inflation as the scalaron, but not the same physics since the equivalence is not complete at the level of the scalar field representations, as will be demonstrated below. This is an important novel point of our current work, pointing to the fact that, although within the {\it vacuumon representation} the RVM can describe primeval inflation, nonetheless inflation is realised in a different physical context than in the Starobinsky approach.


With the above understanding, we now proceed with equating the right-hand sides of Eqs.(\ref{staropotent}) and (\ref{Pott}).
This is understood as valid {\it only} in the appropriate range of the respective scalar fields, depicted in fig.~\ref{fig:potstar2}, given the fact that the potential for the vacuumon \eqref{Pott} is bounded, while the Starobinsky potential is only bounded for $\varphi > 0$.

In this way, we
can express the {\it early vacuumon} field as a function
of the {\it scalaron}. After some algebra, we obtain:
\begin{equation}\label{XXXa}
\phi(\varphi)=\frac{1}{\kappa}{\rm ln}\left[ \chi(\varphi)+\sqrt{\chi(\varphi)^{2}-1}\right]
\end{equation}
where
\begin{equation}
\label{XXX}
\chi(\varphi)=\left[ \frac{1+\sqrt{1+8F(\varphi)}}{2F(\varphi)}\right]^{1/2}~,
\end{equation}
with
\be
F(\varphi)=\frac{3\alpha {\cal M}^2\left( 1 - e^{-\sqrt{\frac{2}{3}} \, \ka\varphi } \right)^2}{4\,H^{2}_{I}} = \frac{\alpha \kappa^{2}}{H^{2}_{I}}V_{\rm eff}(\varphi)>0 \;.
\ee
{We note that $0<F(\varphi)\ll 1$ since $\alpha\ll1$ and the {\it scalaron} mass is of order of the inflationary scale.  As a result, $\chi(\varphi)>1$  and hence the former relation between the { early vacuumon} and the {\it scalaron},  Eq.\,(\ref{XXXa}), is well-defined and leads to $\phi>0$ in the early universe. }

In Fig.~\ref{fig:potstar2} we plot
the RVM effective potential $\alpha \kappa^{2} U/H^{2}_{I}$
(solid curve) versus $\kappa \varphi$. On top of that
we show
the effective Starobinsky (dashed curve) potential, {specifically $4(\kappa^2/{\cal M}^2)\,V_{\rm eff}(\varphi)$.}
Clearly,
although the RVM and Starobinsky's model have
different geometrical origins, namely GR and
$R^{2}$, the potentials for the two models look {\em similar} from the viewpoint
of those properties of inflation that can be extracted by an
effective scalar-field dynamics.

However, as already mentioned, this similarity between the models is confined only to the form of the respective potentials. Apart from the opposite signs with which the two potentials enter the respective effective actions, as discussed previously, below Eq.~\eqref{steps2}, when expressed in terms of the scalaron, the kinetic term of the vacuumon would look non canonical, and this already manifests the important difference between the two models. Moreover, as already mentioned, we cannot consider the standard inflaton fluctuations of the vacuumon, since the latter is treated purely as a classical effective description. The slow roll parameters computed naively from
the potential \eqref{Pott} are 
compatible with Planck data~\cite{Planck} at $\sim 2\sigma$ level.

In this respect, we mention that within the RVM
effective approach, the microscopic origin of the
de Sitter space time, which would lead to an understanding of
the quantum fluctuations, is not specified. Additional input from specific models is necessary for this purposes.
For example, if higher curvature corrections \`a la Starobinsky are present, inflation might be due to the scalaron (quantum) field mode, which fits excellently the data~\cite{Planck}. Alternatively, in the string-inspired scenario of \cite{Anomaly2019a}, primordial gravitational waves are responsible for inducing condensates of the anomaly term, which in turn leads to a de Sitter space time with the $H^4$ contributions to the running vacuum energy density. Quantum fluctuations of the condensate are not equivalent to the scalaron quantum flcutations, and in fact such a computation is pending, although the fluctuations are expected to be strongly suppressed, so that scale invariance should be approximately intact.

We would also like to stress that
the RVM  renders  a simple explanation of both graceful exit and reheating problem and provides a unified view of the cosmic evolution, see
\,\cite{LBS2013,BLS2013,LBS2014,SolaGRF2015,LBS2016} for details.
 The process of reheating after the exit of the inflationary epoch in Starobinsky's model has been studied
e.g. in \cite{ReheatStaro}. However, in contrast to the conventional scenarios, there is no genuine reheating for the vacuumon.  
{There is, instead, a smooth transition from the early vacuum energy into radiation, without intervening
particle decays.  Rather than reheating there is a progressive heating up of the universe during the massive conversion of vacuum energy into radiation. This mechanism triggers a very large entropy production and may  render an alternative solution to the entropy problem, see \cite{SolaGRF2015,LBS2016}.}

\subsection{Equation of state of the early vacuumon}\label{EoS early vacuumon}
The EoS of a given cosmological model can be a useful tool to describe important physical properties of such model.  From the formulas derived in the previous section it is easy to find the following appropriate expression for the effective EoS of our system:
\begin{equation}\label{EOStotal}
w=\frac{p_\phi}{\rho_\phi}=
\frac{\frac12\,\dot{\phi}^2-U(a)}{\frac12\,\dot{\phi}^2+U(a)}=-1-\frac{a}{3H^2}\frac{dH^2}{da}\,,
\end{equation}
where the kinetic term has been written as $(1/2)\dot{\phi}^2=(-1/\kappa^2)\,\dot{H}=(-1/\kappa^2)\,aH\,H'= (-1/2\kappa^2)\,a\,dH^2/da$ and use has been made of the equation (\ref{Vz}) for the effective potential.
Notice that in the early universe we are dealing with a transition from the primeval vacuum energy into a heat bath of radiation, and therefore the EoS of the vacuumon should reflect this transition.  Because matter is essentially relativistic in the early universe,  we have
$\omega_m=1/3$  in Eq.\,(\ref{HS1}).  Using this expression we can compute the EoS of the early vacuumon from (\ref{EOStotal}).  A simple calculation renders
\begin{equation}\label{EoSearlyvacuumon1}
w=-1+\frac43\,\frac{D a^4}{D a^4+1}\,,
\end{equation}
where again we neglected the $H^2$ terms in the early universe in front of the dominant power $H^4$ in Eq.(\ref{lambda}).  The previous equation can be rephrased in a more suggestive way as follows. Let us compute the transition point (call it $a_{*}$) where the vacuum energy density and the radiation energy density become equal. It  ensues from equating equations (\ref{eq:rLa}) and (\ref{eq:rhor}). We find that the coefficient  $D$ becomes determined  in terms of the scale factor at the transition point:
\begin{equation}\label{eq:transition}
  D=\frac{1}{a_*^4}\,.
\end{equation}
As a result the EoS (\ref{EoSearlyvacuumon1}) can be rewritten as follows:
\begin{equation}\label{EoSearlyvacuumon2}
w=-1+\frac43\frac{\left(\displaystyle{\frac{a}{a_*}}\right)^4}{\left(\displaystyle{\frac{a}{a_*}}\right)^4+1}\,,
\end{equation}
Notice that the point $a_*$ represents the nominal end of inflation and the start of the radiation epoch.  In fact, from the previous equation we can easily see that for $a\gg a_*$ (i.e. deep in the radiation epoch) we have $w\simeq 1/3$, whereas for $a\ll a_*$ (i.e. deep in the inflationary epoch) we have $w\simeq -1$.  This behavior confirms our interpretation of the early cosmic evolution in terms of the vacuumon. As we will see in Sect. \ref{EoS late vacuumon}, an EoS analysis (in this case of the ``late vacuumon'') can be particularly enlightening for studies of the properties of the DE in the current universe. In the next section we shall study the late running vacuum universe from this perspective.

\subsection{Scalar field description of the total cosmic fluid in the late universe}\label{LateUniverse}
{In the previous sections  we have seen that the scalar field language helps to show that
the RVM can describe inflation in a successful way, comparable to the scalaron.  In the following two sections we show that it also helps  to describe the current universe, comparable to scalar or phantom DE fields.}
In this section let us again focus on Eq.(\ref{HE}).
As long as the radiation component starts to become sub-dominant the matter
dominated epoch appears. At this point
since the early de Sitter era is left well
behind ($H \ll H_{I}$),
the quantity $c_{0}/H^{2}$ in
Eq.(\ref{HE}) starts to dominate over $\alpha
H^{2}/H_{I}^{2}$. Therefore, Eq.(\ref{lambda}) reduce to
\begin{equation}\label{l0cdm}
\Lambda(H)=\tilde \Lambda_{0}+3\nu(H^{2}-H_{0}^{2}),
\end{equation}
where  $\tilde \Lambda_{0}=3c_{0}+3\nu H^{2}_{0}$
is the current value of vacuum (cosmological
constant) energy density.
Using the operator $d/dt=aH\,d/da$ and
taking into account that after recombination the cosmic fluid consists
dust ($\omega_m=0$) and running vacuum
with $H\ll H_I$, we can rewrite Eq.(\ref{HE}) as follows
$$
a\,\frac{dH^2}{da}+3(1-\nu)H^2-3\,c_0=0\,.
$$
Therefore, the corresponding
solution obeying the boundary condition $H=H_0$ at the present
time ($a=1$) is:
\be
\label{Hz}
H^{2}(a) = H_0^2
\left[{\tilde \Omega}_{m0}\,a^{-3(1-\nu)}+{\tilde \Omega}_{\Lambda 0}\right],
\ee
with
$$
{\tilde \Omega_{\Lambda 0}}=1-{\tilde \Omega_{m0}}=\frac{\Omega_{\Lambda 0}-\nu}{1-\nu},
$$
where $\Omega_{\Lambda 0}=1-\Omega_{m0}$.
The above boundary condition fixes the value of
the parameter $c_0$ as follows: $c_0=H_0^2(\Omega_{\Lambda 0}-\nu)$.
Here the matter and vacuum densities are given by
\be\label{RVMrhom}
\rho_{m}(a)=\rho_{m0}a^{-3(1-\nu)}
\ee
\be\label{RVMrhoL}
\rho_{\Lambda}(a)=\rho_{\Lambda 0}+\frac{\nu \rho_{m0}}{1-\nu}[a^{-3(1-\nu)}-1].
\ee
Of course in the case of
$\nu=0$ we fully recover the concordance $\Lambda$CDM.
It is interesting to mention that even
for small values of $\nu$ the Universe contains
a mildly evolving vacuum energy that could appear as
dynamical dark energy.
It has been found that the current cosmological model
is in excellent agreement
with the latest expansion data and
it provides a growth rate of clustering which is
compatible with
the observations (for more details
\cite{GoSolBas2015,BPS09,GoSol2015,GrandeET11}).


Furthermore, substituting
Eq. (\ref{Hz}) as well as its
derivative $(H^{'})$ into Eq. (\ref{ppz}) we obtain
\begin{equation}\label{phia1}
\phi(a)=-\frac{\sqrt{3(1-\nu){\tilde \Omega}_{m}}}{\kappa}\int
\frac{da}{a\left[{\tilde \Omega}_{m0} +{\tilde \Omega}_{\Lambda0} a^{3(1-\nu)}\right]^{1/2}},
\end{equation}
where we have used the minus sign here in order to ensure
continuity of the effective potential. At this point it is important
to notice that in order to derive Eq. (\ref{phia1})
we have set
$\rho_{\rm tot}\equiv \rho_{m}+\rho_{\Lambda}={\dot \phi}^{2}/2+U(\phi)$
and $p_{\rm tot}\equiv p_{m}+p_{\Lambda}={\dot \phi}^{2}/2-U(\phi)$.
This implies that the current scalar field description refers to
the total cosmic fluid, hence
the corresponding effective equation of state parameter
is given by
$$
w_{\rm tot}=\frac{p_{m}+p_{\Lambda}}{\rho_{m}+\rho_{\Lambda}}=
\frac{\frac12\,\dot{\phi}^2-U(a)}{\frac12\,\dot{\phi}^2+U(a)},
$$
where for non-relativistic matter we have $p_{m}=0$.
Notice that here we are well inside the matter (non-relativistic) era,
and thus the radiation contribution to the cosmic expansion is negligible.
In the next section we will introduce another dynamical field (we shall call it $\chi$)
as an effective description of $\Lambda(H)$ in the late Universe.

Changing the integration variable as follows
\begin{equation}\label{phia11}
a=\left( \frac{{\tilde \Omega}_{m0}}{{\tilde \Omega}_{\Lambda0}}\right)^{\frac{1}{3(1-\nu)}}
{\rm sinh}^{\frac{2}{3(1-\nu)}}(u)
\end{equation}
{the integral can can be performed analytically. Taking the range $(0, a)$ for the scale factor, which corresponds
to $(0,u)$ in the transformed variable, we find}
\be\label{phia111}
\phi=A\;\ln\left| \frac{\sinh u}{1+\cosh u}\right|=A\;{\rm ln}\left|\frac{e^{u}-1}{e^{u}+1}\right|\,,
\ee
{
where $A = -2/\kappa \sqrt{3(1-\nu)}$.
With the aid of Eq.(\ref{phia11})
the evolution of the { late vacuumon} field can be found explicitly in terms of the scale factor:}


\be \label{phia2}
\phi(a)=A\ {\rm ln} \left[ \frac{ \sqrt{ r a^{3(1-\nu)}+1 } +\sqrt{r a^{3(1-\nu)} }-1  }
{ \sqrt{ r a^{3(1-\nu)}+1 } +\sqrt{r a^{3(1-\nu)} }+1  }
\right]\,.
\ee

\begin{figure}
\includegraphics[width=0.45\textwidth]{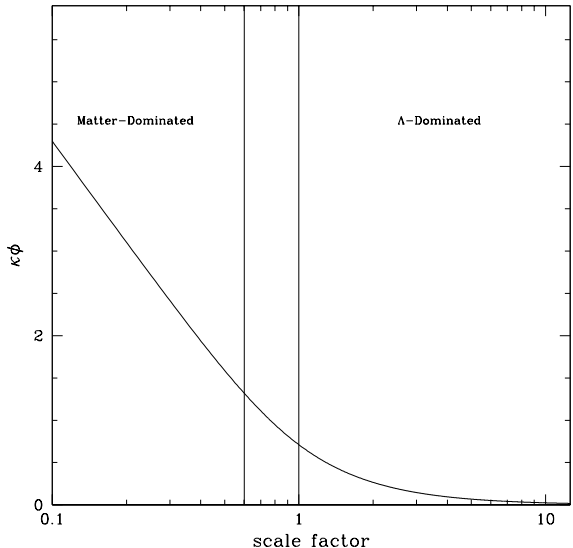}
\caption{The evolution of the late $\phi$ 
field. The area between the
two vertical lines corresponds to that era for which the cosmic
fluid consists of  running vacuum and matter. Here we use
$(\Omega_{m 0},\nu)=(0.30,10^{-3})$.}
\label{fig:potstar2b}
\end{figure}

{where $r$ is the ratio}
\begin{equation}\label{ratior}
r=\frac{\tilde{\Omega}_{\Lambda0}}{\tilde{\Omega}_{m0}}=\frac{\Omega_{\Lambda0}-\nu}{\Omega_{m0}}\,.
\end{equation}
{Such ratio coincides very approximately with the current ratio of vacuum energy density to matter energy density since $|\nu|\ll\Omega_{\Lambda0}$.}
In Fig.~\ref{fig:potstar2b} we provide the evolution of the vacuumon field.
As far as $\Omega_{m 0}$ and $\nu$ are concerned we utilize
the values $(\Omega_{m 0},\nu)=(0.30,10^{-3})$ which are in agreement
with the recent analyses\,\cite{SOLL17,SOLL18,Tsiapi}.

Let us note from Eq.\,(\ref{phia2}) that
$\phi\to +\infty$ for $a\to 0$ (i.e. deep in the past),
whereas  $\phi\to 0$ for $a\to \infty$ (in the remote future).
In actual fact, this result applies strictly only to the late scalar
field , i.e. whenever $H^4$ can be neglected in front of $H^2$ in Eq.\,(\ref{lambda}). In practice this means well after the inflationary epoch, so it comprises most of the radiation epoch and the entire matter-dominated epoch and the DE epoch.
We have not studied the interpolation between the early and late 
regimes here, so the two types of fields 
behave as we have
described only in the mentioned periods.

\begin{figure}
\includegraphics[width=0.45\textwidth]{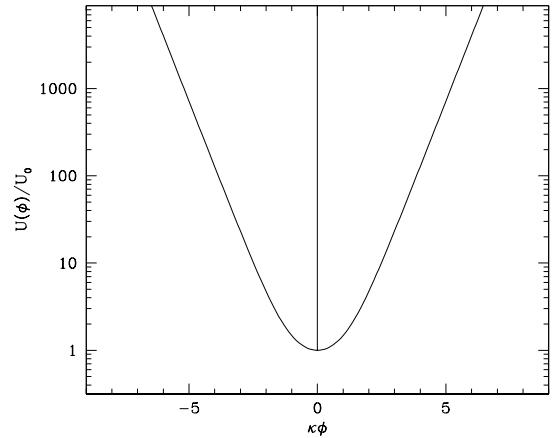}
\caption{The effective potential
$U/U_{0}$ versus the late vacuumon field $\kappa \phi$.
In order to produce the line we use
$\nu=10^{-3}$. The potential has a minimum $U_{\rm min}=U_{0}=
\frac{3H_0^{2}{\tilde \Omega}_{\Lambda 0} } {\kappa^{2}}$
at $\phi=0$ which corresponds to the far future ($a\gg 1$), corresponding
to the late de Sitter era.  Despite we show the symmetric
form of the potential around  $\phi=0$,  by definition
the late scalar 
field of the total cosmic fluid can not take negative values. See section IVC for details.}
\label{fig:potstar3b}
\end{figure}

Concerning the evolution of the potential, utilizing
Eqs. (\ref{Hz}) and (\ref{Vz}) we easily find that
\begin{equation}\label{vdea}
U(a) = \frac{3H_0^{2}}{\kappa^{2}} \left[{\tilde \Omega}_{\Lambda 0} +
\frac{1+\nu}{2}
{\tilde \Omega}_{m0}a^{-3(1-\nu)}\right]\;.
\end{equation}
Notice that in the far future $a\gg 1$ the potential tends to
a constant value
$U\to \frac{3H_0^{2}{\tilde \Omega}_{\Lambda 0} } {\kappa^{2}}$, which
means that the universe enters the
final de Sitter era \cite{LBS2013,Perico2013}. {This is perfectly consistent with the constant asymptotic behavior of the running
vacuum energy density (\ref{RVMrhoL}).}
At this point we attempt to write the
potential in terms of the 
total field $\phi$.
Inserting Eq.(\ref{phia11}) into Eq.(\ref{vdea}) the potential is written as
\begin{equation}\label{vdea1}
U(u) = \frac{3H_0^{2}{\tilde \Omega}_{\Lambda 0} } {\kappa^{2}}
\left[1 +
\frac{1+\nu}{2 {\sinh}^{2}(u)}\right]\;.
\end{equation}
Inverting Eq.(\ref{phia111}), namely
$e^{u}=\frac{1+e^{\omega \phi}}{1-e^{\omega \phi}}$
and using the definition
${\rm \sinh}(u)=\frac{e^{u}-e^{-u}}{2}$, it is easy to prove that
\be\label{sinhsinh}
 {\rm sinh}(u)=-\frac{1}{{\rm sinh}(\omega \phi)}\,.
 \ee
 {This equation shows once more that $\phi$ cannot take positive values as $\sinh u$ must always be positive
 in order to have a well defined scale factor, see Eq.\,(\ref{phia11}).    Equation (\ref{sinhsinh})
 implies}
\begin{equation}\label{vdea1b}
U(\phi) = U_{0}
\left[1 + \frac{1+\nu}{2} {\rm sinh}^{2}(\omega \phi)\right]\;,
\end{equation}
where $\omega=1/A$ and
$U_{0}=U(0)=\frac{3H_0^{2}{\tilde \Omega}_{\Lambda 0} } {\kappa^{2}}$.
Obviously, the potential has a minimum ($U_{\rm min}=U_{0}$) at $\phi=0$
which corresponds to $a\gg 1$.

In order to visualize the RVM effective
potential $U(\phi)$ as a function of $\phi$
in Fig.~\ref{fig:potstar3b} we plot
the ratio $U/U_{0}$ at late times
as a function of $\kappa \phi$.
{Note that although the potential (\ref{vdea1}) is an even function of $\phi$, only the branch $\phi\leq0$ is physically meaningful, as we explained before. The evolution of the universe terminates at the infinite future, corresponding to $\phi=0$.}
{As expected, the limit $\nu\to0$
of all the formulas in this section corresponds to the $\CC$CDM. While the deviation of the RVM and of its effective scalar description from
the $\CC$CDM is, of course, small at the present time (as shown by the value $\nu\sim 10^{-3}$ preferred by the
current fits to the data), the departure of the vacuum energy density from a strict constant is exactly the reason why a mild dynamical dark energy behavior is possible at present
and is also responsible for the improved fits as compared to the $\CC$CDM\,. \cite{SOLL17,SOLL18,Tsiapi}. The theoretical reason for such effect has been explained in detail in\,\cite{AdriaJoan2017}. }


\subsection{Equation of state of the late vacuumon and the phantom effective EoS}\label{EoS late vacuumon}
Here we repeat the analysis of the EoS made in Sect.\,\ref{EoS early vacuumon}, but now for the late vacuumon, which is specially pertinent since it is sensitive to the features of the DE at present, and hence potentially measurable.  From the equations of the previous section and the general equation (\ref{EOStotal}), along with Eq.\,(\ref{Hz}), we find
\begin{equation}\label{EOStotal2}
w=-1+(1-\nu)\tilde{\Omega}_m\frac{H_0^2}{H^2}\,a^{-3(1-\nu)}\,.
\end{equation}
The latter can be worked out as follows:
\begin{equation}\label{EOStotal3}
w(a)=-1+\frac{1-\nu}{1+r\,a^{3(1-\nu)}}\,.
\end{equation}
where $r$ is the ratio defined before in Eq.\,(\ref{ratior}).
The above formula (\ref{EOStotal3})  shows in a transparent way the physical behavior: deep in the matter-dominated epoch ($a\ll1$) the EoS tends to a very small value $w\to -\nu$ (such value would be exactly zero in the $\CC$CDM), whereas deep in the DE epoch ($a\gg 1$) the EoS $w\to -1$.
This kind of evolution was expected since the late vacuumon field describes the combined system of matter and vacuum energy. Therefore the EoS (\ref{EOStotal3}) of the compound system transits from a situation of matter dominance ($w\simeq 0$) into a future one of vacuum dominance ($w=-1$). This is similar to the role played by the early vacuumon, which described the transition from the epoch of inflation into the radiation dominated epoch.  At present we find ourselves in a mixed EoS state of nonrelativistic matter and DE. Let us expand formula (\ref{EOStotal3}) linearly in $\nu$ in order to identify what is the leading contribution from the vacuum dynamics to the EoS around the present time. We find:
\begin{equation}\label{wexpanded}
w(a) \simeq -1+\frac{1-\nu}{1+r\,a^3}\,\left(1+3\nu\, r\,\frac{a^3\,\ln a}{1+r\,a^3}\right)\,.
\end{equation}
The terms of ${\cal O}(\nu)$ give the leading correction introduced by the vacuum dynamics.  However, in this formulation  matter and vacuum are in interaction and one cannot disentangle one from another, in particular we cannot read off the effective EoS of the vacuum. As indicated above, only  in the asymptotic regime the pure matter or vacuum EoS's  are recovered in opposite ends of the cosmological evolution. On the other hand, while the universe is in transit between the remote past and the remote future, the vacuumon field is evolving in a nontrivial way (see Eq.\,(\ref{phia2}) and Fig.\,2). Hence, the vacuumon field can describe the vacuum state when it is near these two ends of the Universe's history, but in  the intervening period its EoS is a mixed one, as we have seen above.

The \textit{vacuumon representation}, therefore, proves to be particularly useful for a description of the physics of the early universe during inflation and at the first stages of the transition into the radiation dominated epoch. During those eras, it provides a sufficiently faithful description of the vacuum evolution, with the rapid inflation period being triggered by the $\sim H^4$ term in the RVM energy density (\ref{lambda}).  In this period, the vacuumon might naively be thought of as effectively playing the  r\^ole of the inflaton, but, as we have mentioned previously, it is quite different from a traditional inflaton field in that it is a classical field, {\it not} a 
fully fledged quantum field degree of freedom and hence it 
does not decay into massive particles (that is to say, there is no conventional reheating mechanism). For this reason, the vacuumon is distinct from the scalaron of ``$R^2$-driven inflation'' and furnishes a different mechanism of inflation \,\cite{LBS2016}.  The reader should recall from Sec. \ref{deSiter} that this was confirmed on formal grounds through a mapping between the scalar field potentials of the two models, which, however, is {\it not} extendable to the complete Lagrangians underlying the two formulations.
Nonetheless, the two inflationary mechanisms are equally efficient and can both implement graceful exit\,\,\cite{LBS2013,BLS2013,Perico2013,LBS2014,SolaGRF2015,LBS2016}.  Only through more detailed analysis and subsequent confrontation with the CMB data it will be possible to distinguish between these two frameworks.  As noted previously, in  both cases a fundamental theory underlies those frameworks: in the RVM case, the form of the vacuum energy density is naturally triggered by the CP-violating gravitational anomalous (Chern Simons) terms that characterise the effective action of the bosonic gravitational multiplet of string theory in a de Sitter background\,\cite{Anomaly2019a,GRF2019}, whereas the scalaron is long known to be associated with the traditional Starobinsky type of inflation\, linked to the conformal anomaly\cite{staro,staro2}.

{Let us now focus on the late universe. We remind the reader
that in the previous section we have introduced the field $\phi$ in
order to describe the total cosmic fluid, namely matter and $\Lambda(H)$
in the late universe.
Here the vacuumon EoS becomes entangled with that of nonrelativistic matter. Hence, some strategy must be devised to track the vacuum evolution. Notice that,  since the EoS of the vacuum is always given by Eq.\,(\ref{eosrvm}) and the EoS of the vacuumon becomes now a time-evolving mixture, we need a different strategy to isolate the genuine effects of the vacuum dynamics in the current era. To this end, an alternative scalar field formulation of the combined system of matter and dynamical vacuum proves convenient.  Despite the fact that $\CC$CDM has exactly the same EoS (\ref{eosrvm}) as the running vacuum, the rigid nature of the (constant in time) vacuum energy density would make it impossible to mimic any form of DE other than the pure vacuum one. On the contrary, in the RVM framework, the time dependence on both sides of Eq.\, (\ref{eosrvm}), which remains valid at all times, enables one to introduce  a dynamical scalar field $\chi$ as an effective description of the RVM in the late Universe. This is more suitable for tracking the DE effects near our time. In such an alternative representation, it is natural to assume the absence of any interaction of $\chi$ with matter, in accordance to the usual minimal assumption made in phenomenological studies of possible dynamical DE effects on the observational data. The $\chi$ field, being self-conserved, satisfies}
\be\label{conservationchi}
\dot{\rho}_\chi+3\,(1+w_\chi)\,H\,\,\rho_\chi=0\,.
\ee
The nontrivial character of $\chi$ now resides in the dynamical form of the EoS as a function of the scale factor or redshift,  $w_\chi=w_\chi(a)$, which will be computed below.

Before doing that, we feel stressing once mote that, while the vacuumon $\phi$  describes the entire system of matter and dynamical vacuum in mutual interaction, $\chi$ specifically describes the dynamics of vacuum independent of matter and tracks possible dynamical DE effects from the departure of $w_\chi$ from $-1$.
As already noted,  despite the fact that the underlying RVM EoS is still Eq.\, (\ref{eosrvm}), this strategy allows us to
obtain the same cosmological system as in the case of a noninteracting quintessence or phantom DE field together with locally conserved matter\,\cite{conpr5, BasSol2013,SolStef05}.

The solution of Eq.\,(\ref{conservationchi}) in terms of the scale factor is
\begin{equation}\label{solrho}
\rho_\chi(a)=\rho_{\chi 0}\,\exp\left\{3\,\int_a^1\,da'\,\frac{1+w_\chi(a')}{a'}\right\}\,.
\end{equation}
It follows from this expression  that
\begin{equation}\label{EoSchi1}
  w_\chi(a)=-1-\frac{a}{3}\,\frac{1}{\rho_\chi(a)}\frac{d \rho_\chi(a)}{da}\,.
\end{equation}
To compute the EoS of $\chi$ such that it mimics the running vacuum, let us first note that the Hubble function $H(a)$  of the model in which the DE is represented by $\rho_\chi$ obviously satisfies
\begin{equation}\label{H2zchi}
\rho_{\chi}(a)=\,\frac{3H^2(a)}{8\pi G} -\rho_{m0}\,a^{-3}\,.
\end{equation}
 Since the expansion history of the RVM is to be matched by that of the scalar field cosmology based on $\chi$,  we can insert  $H^2(a)$ from the RVM model, Eq.\,(\ref{Hz}), in the previous expression and compute explicitly the derivative involved in (\ref{EoSchi1}).  This yields the effective EoS of $\chi$ that matches the running vacuum. After some calculations one finds that
 \begin{equation}\label{EoSchiFULL1}
w_\chi(a)=-1+(1-\nu)\,\frac{f(a)}
{g(a)}\,,
\end{equation}
where
\begin{equation}\label{EoSchiFULL2}
f(a)=\Omega_{m0}\left(a^{-3(1-\nu)}-a^{-3}\right)
\end{equation}
and
\begin{equation}\label{EoSchiFULL3}
g(a)=\Omega_{m0}^0\,[a^{-3(1-\nu)}-1]-(1-\nu)\,[\Omega_{m0}\,a^{-3}-1]\,.
\end{equation}
If we expand straightforwardly these expressions  linearly in $\nu$ and reexpress the result in terms of the redshift variable $z=a^{-1}-1$, which is more convenient for observations, we arrive at the leading form of the desired  EoS:
\begin{equation}\label{wchiexpanded}
w_\chi (z)\simeq -1- 3\nu\,\frac{\Omega_{m 0}}{\Omega_{\Lambda 0}}\,\,(1+z)^3\,\ln (1+z)\,.
\end{equation}
The departure of the above EoS from $-1$ precisely captures the dynamical vacuum effects of the RVM in the language of quintessence and phantom DE models.
As we can see very obviously from (\ref{wchiexpanded}), the effective behavior is quintessence-like if $\nu<0$ , or  phantom-like if  $\nu>0$.  As an example, let us take the recent fitting results of the RVM from Ref.\,\cite{SOLL18} (cf. Table 1 of this reference).  If we take three redshift points near the transition between deceleration and acceleration, e.g. $z=0.7,  1,  1.5$,  we find $w_{\chi}\simeq-1.005,  -1.016,  -1.030$, respectively, hence a mild phantom-like behavior. As it is well-known, phantom behavior of the DE is perfectly compatible with the observational data\,\cite{Planck} and it has been a bit controversial if interpreted in terms of  fundamental scalar fields, even if playing around with more than one field.

In contrast, the effective description of vacuum in interaction with matter as performed here shows that one can mimic quintessence or phantom DE in models without fundamental fields of this sort. In particular, a phantom-like behavior in our context is completely innocuous since the underlying model is not a fundamental scalar field model. In the current  study the underlying model is the running vacuum model, which, as we noted,  grants improved fits to the overall observations as compared to the $\CC$CDM\,\,\cite{SOLL17,SOLL18,Tsiapi}.  Therefore, the mere observation of quintessence or phantom DE need not be associated to fundamental fields of this kind, it could be the effective behavior of a nontrivial (and phenomenologically successful)  theory of vacuum.
\vspace{0.5cm}

\section{Conclusions \label{sec:5}}
In this article we have provided for the first time
a {\it classical} scalar field description
of the running vacuum model (RVM) {throughout the entire
cosmic evolution}. That model is
based on the renormalization-group approach in curved spacetime.
Specifically, we have found that the RVM  can be described
with the aid of an effective {\it classical} scalar field that we called the
{\it vacuumon}. At early enough times the $H^4$-term of the RVM
dominates the form of running vacuum and it is responsible for
inflation. Such $H^4$-driven mechanism for inflation is
typical of string theories
when their effective action (which contains the gravitational Chern-Simons term) is  averaged over the
 inflationary spacetime,
in the presence of primordial gravitational waves. Within this framework,
the effective {\it early vacuumon} field
can be used towards describing the cosmic expansion,
namely the universe starts from a non-singular initial
de Sitter vacuum stage and it smoothly passes from
inflation to radiation epoch, hence graceful exit is achieved.
Also, we have shown that under certain conditions the {\it early vacuumon}
potential can be made formally equivalent to the
Starobinsky potential, despite the fact that the origins
of the two models (RVM and Starobinsky) are quite different.
However, the two frameworks, namely $H^4$-driven  inflation and $R^2$-driven inflation,
 are, however, not physically equivalent, but both provide a successful description of inflation with graceful exit
 and can be parametrized in terms of scalar fields, the vacuumon and the scalaron, respectively. Nonetheless, the classical nature of the vacuumon does not allow cosmological perturbations to be discussed, unlike the inflaton case. One needs to consider the underlying microscopic models, e.g. string theory as in \cite{Anomaly2019a}, for this.

{If we focus on the low energy physics, the RVM has also been shown
to fit the current observational data rather successfully as compared to the $\CC$CDM,
and here we have also presented an effective scalar field description.}
 The late time cosmic
expansion is still dominated by the constant additive term of the running vacuum density
$\rho_{\Lambda}(H)$, what makes the RVM  to depart only slightly from the $\CC$CDM. However,
the power dependence  $H^2$  in
$\rho_{\Lambda}(H)=\rho_{\Lambda}^0+(3\,\nu/\kappa^2)(H^2-H_0^2)$ makes the current vacuum
slightly dynamical, and this is the clue for some improvement of the RVM fit to the data as compared to the $\CC$CDM.
Since the coefficient in front $H^{2}$ is small the evolution of the
vacuum energy density is mild prior to the present time.
The remnant of the RVM at present epoch
is precisely that mild quadratic dynamical behavior of
$\rho_{\Lambda}(H)$ around the present value $\rho_{\Lambda 0}$
which is affected by the coefficient $|\nu|\ll1$.
The current dark energy dominated era
can also be described in the framework of an alternative scalar field whose equation of state
can appear in the form of quintessence or phantom dark  energy.  Using the fitting data existing in the literature on the RVM we find
that the phantom option is favored in this representation,
which is perfectly compatible with the current observations.
{The overall picture of the universe is that the vacuum
is dominant in the early stages, it decays into radiation, proceeds into matter era
and enters the current epoch of mild dynamical DE until the universe whimpers into
a final de Sitter stage.  The difference with the $\CC$CDM is that here the vacuum is
dynamical at all stages of the evolution, and this also helps in a better description of the data.}

To conclude, in this work we have argued that the effective {\it vacuumon} scalar-field representation
of the RVM provides an efficient way for understanding
the underlying mechanism of early inflation, as is the case of $R^2$-inflation, in which
the scalaron plays such  a r\^ole.  However, we have demonstrated that the vacuumon and the scalaron describe
two physically different mechanisms of inflation.  We have also shown how to describe the intervening period between the two asymptotic de Sitter epochs of the Universe, the inflationary and the current one, by means of an alternative scalar field representation, which is able to track the dynamical character of the vacuum through an ostensible effective phantom behavior, very near  $w=-1$ (approaching it from below) around our time.

\vspace{0.5cm} \textbf{Acknowledgments.} SB acknowledges support from
the Research Center for Astronomy of the Academy of Athens in the 
context of the program  ``{\it Tracing the Cosmic Acceleration}''.
The work  of NEM is supported in part by the UK Science and Technology Facilities  research Council (STFC) under the research grant
ST/P000258/1. The work of JS has
been partially supported by projects  FPA2016-76005-C2-1-P (MINECO), 2017-SGR-929 (Generalitat de Catalunya) and MDM-2014-0369 (ICCUB).
NEM also acknowledges a scientific associateship (``\emph{Doctor Vinculado}'') at IFIC-CSIC-Valencia University, Valencia, Spain.



\begin{thebibliography}{99}

\bibitem{Planck2018}
  N.~Aghanim {\it et al.} [Planck 2018 results. VI. Cosmological parameter],
  arXiv:1807.06209 [astro-ph.CO].

\bibitem{Planck2015}
{Planck Collab. 2015, P.A.R. Ade {\it et al.,} Astron. Astrophys. {\bf 594} (2016) A13. }

\bibitem{Planck2013}
{P.A.R. Ade et al. A\&A. \textbf{571},\ A16 (2014)}

\bibitem {dataacc1}S. Perlmutter, et al., Astrophys. J. \textbf{517}, 565 (1998)

\bibitem {dataacc2}A. G. Riess, et al., Astron J. \textbf{116}, 1009 (1998)

\bibitem {data1}P. Astier et al., Astrophys. J. \textbf{659}, 98 (2007)

\bibitem {data2}N. Suzuki et al., Astrophys. J. \textbf{746}, 85 (2012)

\bibitem {data3}E. Komatsu et al., Astrophys. J. Suppl. \textbf{192}, 18 (2011)

\bibitem {conpr1}S. Weinberg, Rev. Mod. Phys.\textbf{ 61,} 1 (1989)

\bibitem{conpr5}
V. Sahni and A. A. Starobinsky, Int. J. Mod. Phys. D., {\bf 9}, 373 (2000)

\bibitem {Pee2003}P. J. Peebles and B. Ratra, Rev. Mod. Phys. \textbf{75,}
559, (2003)

\bibitem {conpr2}T. Padmanabhan, Phys. Rept. \textbf{380,} 235 (2003)

\bibitem{JSPRev2013} J. Sol\`a,
J. Phys. Conf. Ser. {\bf 453} (2013)  012015 [arXiv:1306.1527]


\bibitem {conpr4}A. Padilla, arXiv:1502.05296


\bibitem {conpr3}L. Perivolaropoulos, arXiv:0811.4684

\bibitem{Riess2016and2018}
A.G. Riess {\it et al.,} ApJ {\bf 826} (2016) 56;
ApJ {\bf 855} (2018) 136.

\bibitem{Macaulay2013}
E. Macaulay, I.K. Wehus \& H.C. Eriksen, Phys. Rev. Lett. {\bf 111} (2013) 16130.

\bibitem{SOLL17}
J. Sol\`a, A. G\'omez-Valent and J. de Cruz P\'{e}rez, Astrophys.J. {\bf 811} (2015) L14 [arXiv:1506.05793]; Astrophys. J. \textbf{836} (2017)  43 [arXiv:1602.02103];
Physics Letters B, {\bf 774}  (2017) 317 [arXiv:1705.06723]

\bibitem{SOLL18}
J. Sol\`a, J. de Cruz P\'{e}rez and A. G\'omez-Valent,
MNRAS  {\bf 478}  (2018) 4357 [arXiv:1703.08218];  EPL {\bf 121} (2018) 39001 [arXiv:1606.00450]


\bibitem{JSP17}   J. Sol\`a,   Int. J. Mod. Phys.  A{\bf 33} (2018) 1844009

\bibitem{SolGoCruz18} J. Sol\`a, A. G\'omez-Valent and J. de Cruz P\'{e}rez,  Phys. Dark  Univ. {\bf 25} (2019) 100311 [arXiv:1811.03505] 

\bibitem{Mehdi19}  	M. Rezaei, M. Malekjani and J. Sol\`a,
Phys. Rev. D{\bf 100} (2019)  023539 [arXiv:1905.00100].

\bibitem{BD-RVM} J. Sol\`a, A. G\'omez-Valent, J. de Cruz P\`erez and C. Moreno-Pulido,  Astrophys. J. Lett. 886 (2019) L6 [arXiv:1909.02554], doi:10.3847/2041-8213/ab53e9


\bibitem {hor1}G.W. Horndeski, Int. J. Theor. Phys. \textbf{10, }363 (1974)

\bibitem {hor2}C. Brans and R.H. Dicke, Phys. Rev. \textbf{124,} 925 (1961)

\bibitem {hor3}A. Nicolis, R. Rattazzi and E. Trincherini, Phys. Rev. D
\textbf{79}, 064036 (2009)

\bibitem {hor4}L. Arturo Urena-Lopez, J. Phys. Conf Ser. \textbf{761}, 012076 (2016)

\bibitem{ShapSol}
I.~L. Shapiro and J.~Sol{\`a}, JHEP {\bf 0202} (2002) 006 [hep-th/0012227]; Phys.Lett. {\bf B475} (2000) 236 [hep-ph/9910462]

\bibitem{ShapSol2}
I.~L. Shapiro and J.~Sol{\`a},  Nucl. Phys. Proc. Suppl. 127 (2004) 71;  PoS AHEP2003 (2003) 013 [astro-ph/0401015].


\bibitem{ShapSolStef}  I.~L. Shapiro, J. Sol\`a, H.
\v{S}tefan\v{c}i\'{c} JCAP 0501 (2005) 012 [hep-ph/0410095]

\bibitem{SolStef05}  J. Sol\`a, H. \v{S}tefan\v{c}i\'{c}, {Phys. Lett.}
    {\bf B624} (2005) {147} [ astro-ph/0505133];  Mod. Phys. Lett.
{\bf A21} (2006) 479  [astro-ph/0507110]

   \bibitem{Fossil07}  J.~Sol\`a,
  J.\ Phys.\ A {\bf 41}  (2008) 164066
  [arXiv:0710.4151]


  \bibitem{ShapSol09} I.~L. Shapiro and J.~Sol{\`a}, Phys.\ Lett.\ B {\bf 682}, 105 (2009) [ arXiv:0910.4925];  arXiv:0808.0315

  \bibitem{JSPRevs} J. Sol\`a,
 AIP Conf. Proc.  {\bf 1606} (2014) 19 [arXiv:1402.7049];
J. Phys. Conf. Ser. {\bf 283} (2011) 012033 [ arXiv:1102.1815]

\bibitem{SolGo2015} J. Sol\`a, and A. G\'omez-Valent,
Int. J. of Mod. Phys. {\bf D24} (2015) 1541003 [arXiv:1501.03832]



\bibitem {Ozer}M. Ozer and O. Taha, Phys. Lett. B \textbf{171}, 363 (1986)

\bibitem {bertolami}O. Bertolami, Nuovo Cimento \textbf{93}, 36 (1986)

\bibitem {chenwu}W. Chen and Y.S. Wu, Phys. Rev. D \textbf{41}, 695 (1990)

\bibitem {lim00}J. A. S. Lima and J. C. Carvalho, Gen. Rel. Grav.
\textbf{26},909 (1994)


\bibitem {lima7}J. V. Cunha, J.A. S. Lima and N. Pires, Astron. and Astrophys.
\textbf{390},809 (2002)

\bibitem {many1}M. V. John and K. B. Joseph, Phys. Rev. D \textbf{61}, 087304 (2000)

\bibitem {many2}M. Novello, J. Barcelos-Neto and J. M. Salim, Class. Quant.
Grav. \textbf{18}, 1261 (2001)



\bibitem {aldrovandi}R. Aldrovandi, J. P. Beltran Almeida and J.G. Pereira,
Grav. Cosmol. \textbf{11}, 277 (2005)

\bibitem {Schutz}R. Schutzhold, Phys. Rev. Lett. \textbf{89}, 081302 (2002)

\bibitem {Schutz2}R. Schutzhold, Int. J. Mod. Phys. A \textbf{17}, 4359 (2002)

\bibitem {Lima1}J. C. Carvalho, J. A. S. Lima, and I. Waga, Phys. Rev. D
\textbf{46}, 2404 (1992)

\bibitem {Lima2}J. A. S. Lima and J. M. F. Maia, Mod. Phys. Lett. A
\textbf{08}, 591 (1993)

\bibitem {Lima4}J. A. S. Lima and M. Trodden, Phys. Rev. D \textbf{53}, 4280 (1996)

\bibitem {Lima5}S. Carneiro, J.A.S. Lima, \ Int. J. Mod. Phys. A \textbf{20},
2465 (2005)

\bibitem {SalimWaga}J. Salim and I. Waga, Class. Quant. Grav. \textbf{10},
1767 (1993)

\bibitem {Waga}R. C. Arcuri and I. Waga, Phys. Rev. D \textbf{50}, 2928 (1994)

\bibitem {span1}S. Pan, MPLA \textbf{33}, 1850003 (2018)

\bibitem{LBS2013}
J.~A.~S.~Lima, S.~Basilakos and J.~Sol\`a,
  MNRAS  {\bf 431}, 923 (2013).

\bibitem{LBS2014} J.A.S. Lima, S, Basilakos, J. Sol\`a,
 Gen. Rel. Grav. {\bf 47} (2015)  40 [arXiv:1412.5196].

\bibitem{SolaGRF2015} J. Sol\`a,  Int. J. Mod. Phys. D{\bf 24} (2015) 1544027; J. Sol\`a and H. Yu,  arXiv:1910.01638.

\bibitem{LBS2016}  
J.A.S. Lima, S. Basilakos, J. Sol\`a,  Eur. Phys. J. C{\bf 76} (2016) 228 [arXiv:1509.00163].


\bibitem{BLS2013}  S.~Basilakos, J.~A.~S.~Lima and J.~Sol\`a,
  Int.\ J.\ Mod.\ Phys.\ D {\bf 22}, 1342008 (2013);
  Int.J.Mod.Phys. {\bf D23} (2014) 1442011.


\bibitem{Perico2013}
  E.~L.~D.~Perico, J.~A.~S.~Lima, S.~Basilakos and J.~Sol\`a,
  Phys.\ Rev.\ D {\bf 88}, 063531 (2013).




\bibitem{GoSolBas2015} A. G\'omez-Valent, J. Sol\`{a}, S. Basilakos,  JCAP {\bf  01} (2015) 004 [arXiv:1409.7048]

\bibitem{GoSol2015}
   A. G\'omez-Valent, J. Sol\`{a}, MNRAS  {\bf 448} (2015) 2810 [arXiv:1412.3785]



\bibitem{Elahe2015} A. G\'omez-Valent, E. Karimkhani, J. Sol\`{a},  JCAP {\bf 12} (2015) 048 [arXiv:1509.03298]

\bibitem{BPS09} S.~Basilakos, M.~Plionis, and J.~Sol{\`a},
Phys. Rev. {\bf D80} (2009) 083511 [arXiv:0907.4555]

\bibitem{GrandeET11} J.~Grande, J.~Sol{\`a}, S.~Basilakos and M.~Plionis,
 JCAP {\bf 08}, 007 (2011) [arXiv:1103.4632]

\bibitem{FritzschSola2012}
H. Fritzsch, J. Sol\`a, Class. Quant. Grav.
    {\bf 29} (2012) 215002 [arXiv:1202.5097];
    J. Sol\`a, Int. J. Mod. Phys. {\bf A29} (2014) 1444016 [arXiv:1408.4427]



\bibitem{BasPolSol2012} S.~Basilakos, D.~Polarski, and J.~Sol{\`a},
 Phys. Rev. {\bf D86} (2012) 043010  [arXiv:1204.4806].

\bibitem{BasSol2013}
  S.~Basilakos and J.~Sol{\`a},
  MNRAS {\bf 437} (2014)  3331;
   Phys.Rev. {\bf D90} (2014) 023008.

\bibitem{Cristina} C. Espa\~na-Bonet \textit{et al.}, JCAP {\bf 0402} (2004) 006 [hep-ph/0311171]; Phys.Lett. {\bf B574} (2003) 149 [astro-ph/0303306].

\bibitem{OldPerturb}
J. C. Fabris, I. L. Shapiro and J. Sol\`a, JCAP
{\bf 0702}, 016, (2007) [gr-qc/0609017];
J. Grande et al.,
Class. Quant. Grav., {\bf 27}, 105004 (2010) [arXiv:1001.0259]




\bibitem{BasMavSol15} S.~Basilakos, N.~E.~Mavromatos and J.~Solà,
  Universe {\bf 2} (2016)  14
  [arXiv:1505.04434].



\bibitem{JSPReview16}
J. Sol\`a,   Int. J. Mod. Phys.  A{\bf 31} (2016)  1630035 [arXiv:1612.02449]

\bibitem{MG15}  J. Sol\`a,  A. G\'omez-Valent and J. de Cruz P\'{e}rez,
\textit{Signs of Interacting Vacuum and Dark Matter in the Universe}, Proc. of MG 15, Rome  (to appear),
arXiv:1904.11470

\bibitem{RVM-BD}  	
J. Sol\`a,  Int.J. Mod. Phys. D{\bf 27} (2018) 1847029 [arXiv:1805.09810];
J. de Cruz P\'{e}rez and J. Sol\`a, Mod. Phys. Lett. A{\bf 33} (2018)  1850228 [arXiv:1809.03329]

\bibitem {Overduin98} J.M. Overduin, F.I. Cooperstock, Phys.Rev. D58 (1998) 043506; R. G. Vishwakarma, Class. Quant. Grav. \textbf{18}, 1159 (2001)



\bibitem{ahm}
  J.~Alexandre, N.~Houston and N.~E.~Mavromatos,
  Phys.\ Rev.\ D {\bf 88}, 125017 (2013).
  Int.\ J.\ Mod.\ Phys.\ D {\bf 24}  (2015) 1541004.

\bibitem{ahmstaro}   J.~Alexandre, N.~Houston and N.~E.~Mavromatos,
  Phys.\ Rev.\ D {\bf 89} (2014) 027703


\bibitem{emdyno}
J.~Ellis and N.~E.~Mavromatos,
  Phys.\ Rev.\ D {\bf 88} (2013) 085029.

\bibitem{Planck}
Y. Akrami {\it et al.}  [Planck Collaboration],
arXiv:1807.06211

\bibitem{Anomaly2019a}  	
S. Basilakos, N. E. Mavromatos and J. Sol\`a, \textit{Gravitational and Chiral Anomalies in the Running Vacuum Universe and Matter-Antimatter Asymmetry},
arXiv:1907.04890.

\bibitem{GRF2019}  	
S. Basilakos, N. E. Mavromatos and J. Sol\`a,
Int. J. Mod. Phys. {\bf 28} (2019) 1944002
[arXiv:1905.04685]


\bibitem{AdriaJoan2017}
A. G\'omez-Valent \& J. Sol\`a,
MNRAS  {\bf 478} (2018) 126 [arXiv:1801.08501];
EPL  {\bf 120} (2017) 39001 [arXiv:1711.00692].


\bibitem{ncstrings}
  J.~R.~Ellis, N.~E.~Mavromatos and D.~V.~Nanopoulos,
  Gen.\ Rel.\ Grav.\  {\bf 32}, 943 (2000)
  [gr-qc/9810086].

\bibitem{staro}
  A.~A.~Starobinsky,
  Phys.\ Lett.\ B {\bf 91}, 99 (1980)






\bibitem{KolbTurner} E. W. Kolb and M. S. Turner, \textit{The Early Universe},  Addison-Wesley Publishing Company, 1990.

\bibitem{Dowker1976}
J. S. Dowker and R. Critchley, Phys. Rev. D., {\bf 13}, 3224 (1976)

\bibitem{PSW}  	
R.D. Peccei, J. Sol\`a, C. Wetterich,  Phys. Lett. B{\bf 195} (1987) 183


\bibitem{staro2}   A. A. Starobinsky, Proc. of the Second Seminar Quantum
Theory of Gravity (Moscow, 13-15 Oct. 1981), INR Press, Moscow, 1982,
pp. 58-72 (reprinted in: Quantum Gravity, eds. M. A. Markov, P. C. West,
Plenum Publ. Co., New York, 1984, pp. 103-128)

\bibitem{Vilenkin} A. Vilenkin, Phys. Rev.
D{\bf 32}, 2511 (1985)


\bibitem{ShapSol3} I.~L. Shapiro and J.~Sol{\`a},  Phys.Lett. {\bf B}530 (2002) 236 [hep-ph/0104182]

     	
\bibitem{ShapSol4} I.~L. Shapiro and J.~Sol{\`a}, \textit{A Modified Starobinsky's model of inflation: Anomaly induced inflation, SUSY and graceful exit}, proc. of 11th Russsian Int. Conf. on Theoretical and Experimental Problems of General Relativity and Gravitation, and Int. Workshop on Gravity, Strings and Quantum Field Theory (GRG 11), and 10th Int. Conf. on Supersymmetry and Unification of Fundamental Interactions (SUSY 2002),
[hep-ph/0210329]; Russ.Phys.J. {\bf 45} (2002) 727, Izv.Vuz.Fiz. {\bf 2002N7} (2002) 75.


\bibitem{ReheatStaro}  	
D.S. Gorbunov, A.G. Panin,  Phys.Lett. B{\bf 700} (2011) 157;
E.V. Arbuzova, A.D. Dolgov, L. Reverberi, JCAP {\bf 1202} (2012) 049;
H. Motohashi, A. Nishizawa, Phys.Rev. D{\bf 86} (2012) 083514.



\bibitem{Tsiapi}
P. Tsiapi and S. Basilakos,  MNRAS  {\bf 485} (2019) 2505.






















\end{thebibliography}
\end{document}